\documentclass[aps,pra,twocolumn,floatfix,superscriptaddress]{revtex4-1}
\usepackage{float}
\usepackage{graphicx}
\usepackage{indentfirst}
\usepackage{xcolor}
\usepackage{amsmath,amsfonts,amsthm,bm}
\begin{document}
\title{Theoretical study on the electric field effect on magnetism of Pd/Co/Pt thin films} 
\author{Eszter Simon}
\email{Eszter.Simon@cup.uni-muenchen.de}
\affiliation{Department Chemie/Phys. Chemie, Ludwig-Maximilians-Univerit\"{a}t M\"{u}nchen, Butenandtstr. 5-13, D-81377 M\"{u}nchen,
Germany}
\author{Alberto Marmodoro}
\affiliation{Department Chemie/Phys. Chemie, Ludwig-Maximilians-Univerit\"{a}t M\"{u}nchen, Butenandtstr. 5-13, D-81377 M\"{u}nchen,
  Germany}
\affiliation{Present address: Institute of Physics, Czech Academy of Sciences, Cukrovarnicka 10, CZ-162 53 Prague, Czech Republic}
\author{Sergiy Mankovsky}
\affiliation{Department Chemie/Phys. Chemie, Ludwig-Maximilians-Univerit\"{a}t M\"{u}nchen, Butenandtstr. 5-13, D-81377 M\"{u}nchen,
Germany}
\author{Hubert Ebert}
\affiliation{Department Chemie/Phys. Chemie, Ludwig-Maximilians-Univerit\"{a}t M\"{u}nchen, Butenandtstr. 5-13, D-81377 M\"{u}nchen,
Germany}
\date{\today}
\begin{abstract}
Based on first principles calculations we investigate the electronic and magnetic properties of Pt layers in Pd$(001)$/Co/Pt thin film structures exposed to an external electric field. Due to the Co underlayer, the surface Pt layers have induced moments that are modified by an external electric field. The field induced changes can be explained by the modified spin-dependent orbital hybridization that varies non-linearly with the field strength. We calculate the x-ray absorption and the x-ray magnetic circular dichroism spectra for an applied external electric field and examine its impact on the spectra in the Pt layer around the L$_{2}$ and L$_{3}$ edges. We also determine the layer dependent magneto-crystalline anisotropy and show that the anisotropy can be tuned easily in the different layers by the external electric field.
\end{abstract}


\maketitle

\section{Introduction}
The control of the magnetization of a system by an external electric field, which is known as magneto-electric effect,
has been widely investigated during the last years
experimentally as well as theoretically,
due to its potential application in spintronics
\cite{Prinz1660,Brovko_2014,Zhang2009c}.
The magneto-electric effect was investigated in particular for bulk magnetic compounds with non-collinear magnetic structures \cite{Dzyaloshinsky, Moriya,PhysRevLett.95.057205}, magnetic semiconductors \cite{Chiba2008,Yamada1065} and multi-ferroics \cite{Lottermoser2004,PhysRevLett.108.237201,PhysRevLett.103.257601,PhysRevB.85.134428}.
Within these studies, 
the modification of the magnetic properties by an external electric field has been associated with various 
electronic mechanisms, such as the shift of the Fermi level or a change in the charge carrier density. 

Recent studies have reported that an external electric
 field may affect the physical properties of layered
  systems in a very pronounced way \cite{PhysRevLett.120.157203, Obinata2015}.
For example, for thin films of Pd, it was shown that the
 electric field induces a phase transition from the para- to the ferromagnetic state.
This finding could be explained by the Stoner
 instability caused by the applied electric field that
  leads to a change in the occupation of the electronic states and shifting  that way the Fermi level to a position with a high density of states (DOS).
In the case of magnetic systems, an electric field changes their magnetic properties first of all
due to its influence on the spin polarization of the valence electrons. 
Such a manipulation of the magnetic state by the electric field can lead to interesting and important effects concerning possible applications \cite{Hsu2017,Schott2017,Yang2018}. 
Performing first principles calculations, it was demonstrated, that a well defined change in the magnetic moment can be observed in the case of a ferromagnetic free-standing thin  Fe, Ni or Co film, for which the 
 magnetic moments show a linear dependence on the strength of the external electric field \cite{PhysRevLett.101.137201}. 
 In this case the 
electron  populations in the different spin channels are varied and thus the balance of majority- and minority-spin electrons is distorted leading in turn to a change of the spin magnetic moment in the system. Apart from the spin magnetic moment many other magnetic properties may be controlled by an applied electric field as for example the orbital moment and its anisotropy as well as the magnetic anisotropy energy \cite{PhysRevLett.101.137201,PhysRevLett.102.187201,APL_Chiba}. 

In case of the Co/Pt bilayer system
it was demonstrated 
by means of anomalous Hall effect measurements 
that the Curie temperature of the Co layer can be controlled by an electric field  \cite{Chiba2011}. 
For the Curie temperature of the bulk 3d transition metal alloys a Slater-Pauling like behavior
was found from first principles calculations \cite{Takahashi_2007}. 
However,  experimental results on thin films showed that the Curie temperature is increasing with an increasing number of valence electrons and does not follow the Slater-Pauling like behavior \cite{Chiba2011}.
This finding indicates that in the case of thin films -
when compared to the  bulk situation -
 other mechanisms can play an important role for the magnetic properties in the presence of an electric field. 

Paramagnetic metals, such as Pt and Pd, that are close
to the Stoner instability, have substantial  induced moments due to the proximity effect when deposited on magnetic substrate layers. 
In the case of Pd deposited on the  Pt/Co bilayer system,
 it was demonstrated experimentally and theoretically that the induced magnetic moment of the 
 Pd layer can be controlled 
 by an applied electric field \cite{Obinata2015}. 
 The inter-relation between the influence of an
  electric field on the magnetic state and the electronic structure of Pt deposited on a magnetic 
  substrate   was investigated 
  in a recent work by exploiting 
the component-specific x-ray absorption spectroscopy 
(XAS) together with
  the x-ray magnetic circular dichroism (XMCD) \cite{PhysRevLett.120.157203}.

The XMCD is one of the most powerful probe for investigating the magnetization of layered systems
 in an element resolved way \cite{Okabayashi2018}. 
 The XMCD spectra give  for a magnetized sample
 the difference in absorption
 for left and right-circularly polarized x-rays. 
  XMCD spectra are often analyzed on the basis of the XMCD sum rules, which link the integrals of the XAS and  XMCD  spectra  to the  spin and orbital magnetic moments of the absorbing atom
\cite{PhysRevLett.68.1943, PhysRevLett.70.694,PhysRevB.47.597,PhysRevB.51.1282, PhysRevB.66.094413}.  

Motivated by a recent experimental XMCD study by  
 Yamada et al.\  \cite{PhysRevLett.120.157203}, we investigated the electronic and magnetic properties of Pt layers in the  surface film system  Pd(001)/Co/Pt
 in the presence of an external electric field
 by means of first principles calculations. 
 This way, we investigated 
 how the electric field influences
  the electronic states, magnetic moments, 
  XMCD spectra of the Pt layers and the magnetic anisotropy in the case of the considered systems.  

The paper is organized as follows.
 In Sec.~\ref{sec2}, the  computational methods used are briefly sketched while  in Sec.~\ref{sec3} the results are presented and discussed. Finally, in Sec.~\ref{sec4} we summarize our results.
\section{Computational details \label{sec2}}
All calculations were performed within the 
framework of density functional theory, 
relying on the local spin-density approximation (LSDA).
 For the exchange correlation potential the parametrization of Vosko, Wilk and Nusair was used \cite{vosko1980}. 
 The electronic structure is described 
 on the basis of the Dirac equation, accounting for all
 relativistic effects coherently this way.
Electronic states were represented 
by means of the corresponding Green function 
calculated using the 
spin-polarized Korringa-Kohn-Rostoker (KKR) Green function formalism as implemented in the SPR-KKR code \cite{Ebert_2011, Ebert_prog, PhysRevB.52.8807}. 
The potentials were treated on the level of the atomic sphere approximation (ASA) and 
for the self-consistent calculations 
an angular momentum cut-off of $l_{\rm max}=3$ was used. 
All necessary energy integrations have been done
by sampling $32$ points on a semicircle contour in the upper complex energy semi-plane. Furthermore, 
the $\mathbf{k}$-space integration was done using $750$ points in the
irreducible part of two dimensional Brillouin zone.  

We investigated the Pd(001)/Co$_{n}$/Pt$_{m}$ thin film surface system, where $n$ and $m$ denote the number of  Co and Pt layers, respectively. The first considered model system
 consisted of three Pd layers, $n=2$ or $5$  Co layers, $m=2$ Pt layers and three layers of empty spheres embedded between a semi-infinite Pd substrate and a semi-infinite vacuum region. For the second model system
 the number of Pt layers was varied with the system consisting of two Pd, $n=5$ Co layers, $m=1$, $4$ Pt layers and three layers of empty spheres embedded between the 
 semi-infinite Pd and  vacuum regions. For Pd, Co, Pt and empty layers ideal epitaxial growth was assumed on a fcc(001) textured substrate with the  experimentally in-plane lattice constant of Pd, $a_{0}=3.89$ \AA.  Structural relaxations were neglected in all cases. 

From the obtained self-consistent potentials the X-ray absorption coefficients $\mu_{\lambda} (\omega)$ for the photon energy $\hbar\omega$ and polarization $\lambda$ were calculated using the SPR-KKR Green function method on the basis of Fermi's Golden rule \cite{Ebert_1996, Ebert_2011}. The corresponding XMCD signal,
\begin{equation}
\Delta\mu(\omega)= \frac{1}{2}
 \big(\mu_{+} (\omega) - \mu_{-} (\omega) \big) \; ,
\end{equation}
is defined  as the difference in the absorption for left and right
circularly polarized radiation.
The broadening of the
experimental spectra 
was simulated by a Lorentzian broadening function
with a width parameter of 1 eV. In addition to the XMCD spectra, the magnetic anisotropy (MAE) was  obtained by means of magnetic torque calculations \cite{PhysRevB.74.144411, Bornemann2007}.

The effect of a homogeneous external electric field was modeled by a periodic array of point charges
 in the vacuum region that  behave essentially like a charged capacitor plate.  
In the present calculations the array of point charges or capacitor plate, respectively, 
 was placed in the last vacuum layer.
 This set-up leads to a homogeneous electric field of strength,
\begin{equation}
 E = \frac{Q}{a_{0}^{2}\epsilon_{0}},
\end{equation}
where $Q$ is the charge of the capacitor in unit of the electron's charge,
$\epsilon_{0}$ is the permittivity of vacuum and $a_{0}^{2}$ is the area of
the unit cell for the Pd(001) plane.
 Here, the applied electric field is perpendicular to the surface and for the positively charged condensed plate it points 
 from the vacuum towards the surface 
 increasing this way the spill-out of the electrons
 from the Pt layers to the vacuum
  with increasing electric field strength.

\section{Results \label{sec3}}

\subsection{Variation of the thickness of the Co layer}

To investigate the impact of an external electric field on the electronic structure of the Pd(001)/Co$_{n}$/Pt$_{m}$ system, we focus first on Pd(001)/Co$_{2}$/Pt$_{2}$, corresponding essentially to a system composed of  0.5 nm Co and  0.4 nm Pt films,
as studied recently in Ref. \cite{PhysRevLett.120.157203}. These experiments have been accompanied by theoretical work, however, considering in contrast to the present study a $(111)$-oriented surface.
 From the self-consistent calculations we obtained the spin magnetic moments of the layers as a function of the electric field. The spin-magnetic moments of the Co layers  without an external electric field ($E = 0)$ are $m_{\rm Co_{1}}=1.92\, \mu_{B}$ and  $m_{\rm Co_{2}}=1.89 \, \mu_{B}$ for
  Pd(001)/Co$_{2}$/Pt$_{2}$, where the indices of the Co layers start at the Pd/Co interface.
   
To see an impact of the thickness of the ferromagnetic film on the
spin and orbital polarization of the Pt film, the calculations
have been performed also for Pd(001)/Co$_{5}$/Pt$_{2}$.
In this case, the magnetic moments for four Co layers (for $E = 0$) are very
close to each other, $m_{\rm Co_{2}}\sim m_{\rm Co_{3}}\sim m_{\rm
  Co_{4}}\sim m_{\rm Co_{5}}\sim 1.8\,\mu_{B} $, while for the 
Co layer at the Pd/Co interface the magnetic moment is the largest one, $m_{\rm Co_{1}}\sim1.9\,\mu_{B}$. 
For both systems the change of the magnetic moment of the Co layers  
induced by the electric field
is negligible. 

\begin{figure}[htb]
\centering
\includegraphics[width=1.00\columnwidth]{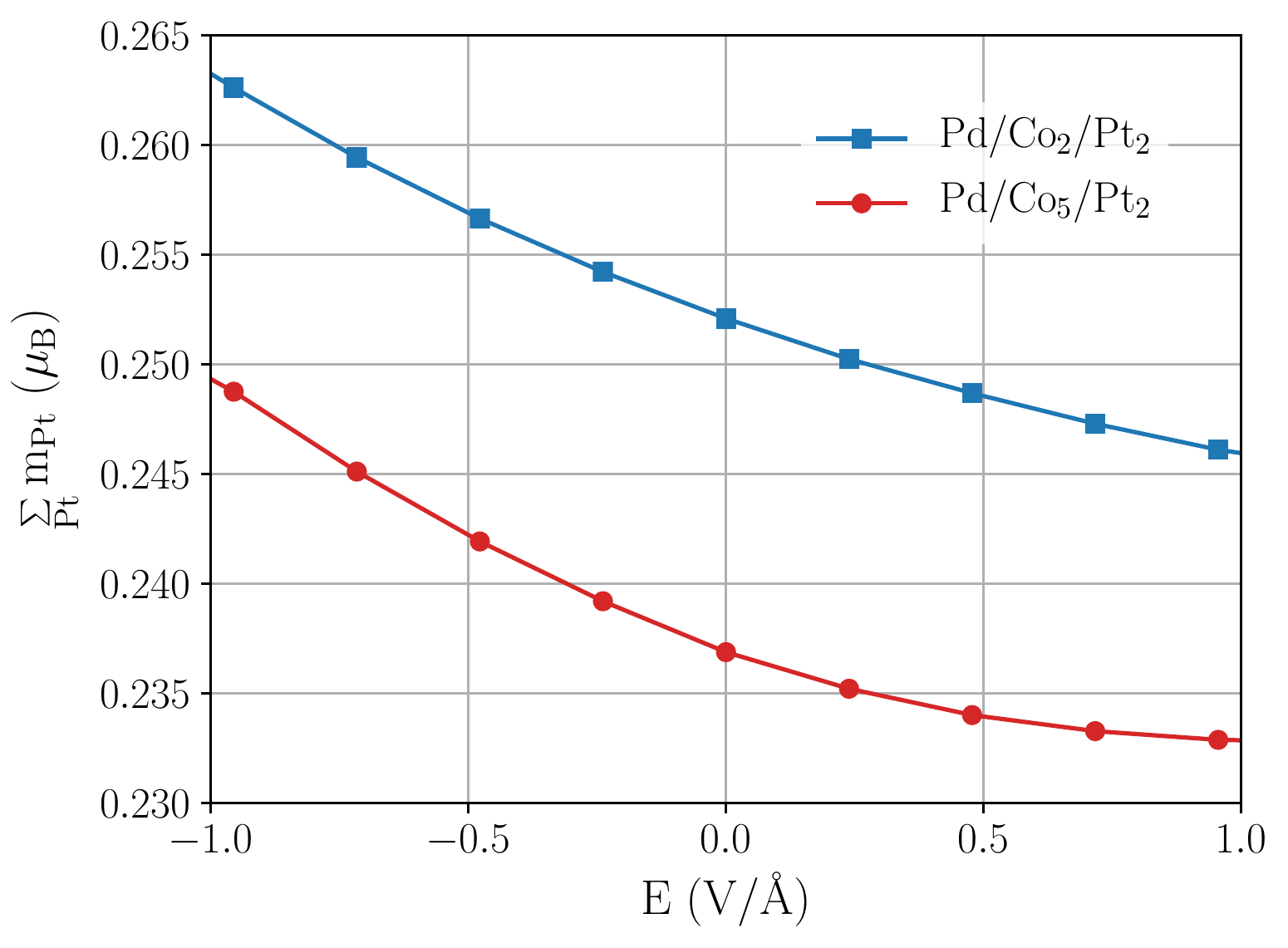}%
\caption{Calculated sum of the spin magnetic moment of
the Pt layers,
$\sum\limits_{\rm Pt} \rm m_{\rm Pt}$, as a function of the external electric field $ E$ for Pd(001)/Co$_{2}$/Pt$_{2}$ (squares) and Pd(001)/Co$_{5}$/Pt$_{2}$ (circles).}
\label{fig_momPt}
\end{figure}

In the Pt layers induced spin moments are formed due to the proximity
effect caused by the Co layer with the value of the induced moment being
largest for the Pt layer at the Pt/Co interface. In the presence of the
external electric field the magnetic moments in the Pt film are
modified. This modification is  
depending on the thickness of the Co film, as can be seen in 
Fig.\ \ref{fig_momPt} 
showing the sum of the spin magnetic moments 
of the Pt layers, $\sum\limits_{\rm Pt} \rm m_{\rm Pt}$, as a function of the electric field for both considered systems.  
%
%
Nevertheless, for both systems the magnetic moment decreases with increasing positive electric field in spite of the different number of Co layers. Moreover, in contrast to calculations for  free standing ferromagnetic thin films \cite{PhysRevLett.101.137201}, the Pt spin-magnetic moment
 in the present case does not vary linearly with the field strength.

\begin{figure}[htb]
\centering
\includegraphics[width=1.00\columnwidth]{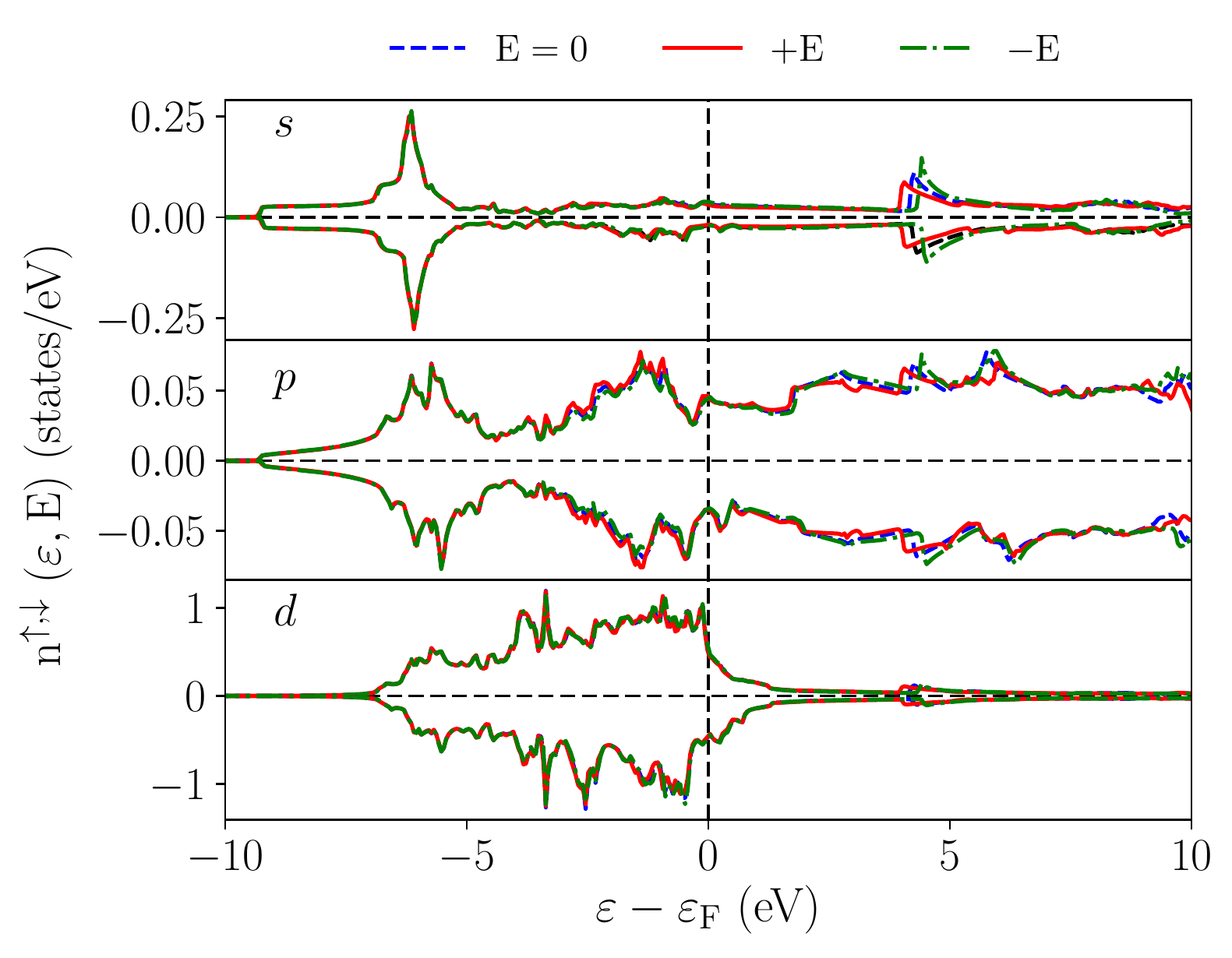}%
\caption{Calculated spin-polarized density of states, 
$n^{\uparrow (\downarrow)}(\varepsilon, \pm E)$, 
projected to $s$, $p$ and $d$ states 
for the topmost Pt layer in 
Pd(001)/Co$_{2}$/Pt$_{2}$ for the  selected electric fields $E = \pm 7$~V/nm
indicated by $\pm E$.}
\label{fig_dosPt}
\end{figure}

In order to understand the impact of an external electric field on the electronic structure of the Pt layer we calculated  the density of states (DOS) for the systems in the presence of the electric field. 
Figure \ref{fig_dosPt}
 shows for various electric field
 strengths
the spin resolved DOS
projected on to $s$, $p$ and $d$ states
for the  topmost Pt layer   in 
 Pd(001)/Co$_{2}$/Pt$_{2}$. 
%
%
The applied electric field strengths have the value $\pm 7$~V/nm denoted as $+ E$ and $- E$ in the following. One can see, that the external electric field slightly modifies the electronic states of the Pt layer. 
A positive electric field shifts the $s$, $p$ as well as $d$ states of Pt down, while a negative field shifts the states up in energy. Moreover, one can see that these shifts increase with the energy of the electronic states as they are more affected by the electric field due to their weaker localization. Due to the difference of the DOS for the two spin channels, these shifts lead to a change of the magnetic moments, which depend  on the sign of the electric field. It should be noted that in contrast to the work on a free standing film in Ref.\ \onlinecite{PhysRevLett.120.157203},  the Fermi level is fixed in our case as we deal with a half-infinite substrate. Accordingly, the value of the Fermi energy is that of the Pd substrate for all applied electric fields. Another effect of the electric field seen in
 Fig.\ \ref{fig_dosPt} 
 is the change of the amplitude of the DOS due to the change of hybridization of the electronic states of Pt and Co layers, or as it was pointed out in Ref.\ \onlinecite{PhysRevLett.120.157203}, the hybridization of the $sp$-states and the $d$-states of Pt. As the hybridization is spin dependent, this also results in a change of the magnetic moments induced by the electric field.

\begin{figure}[htb]
\centering
\includegraphics[width=1.00\columnwidth]{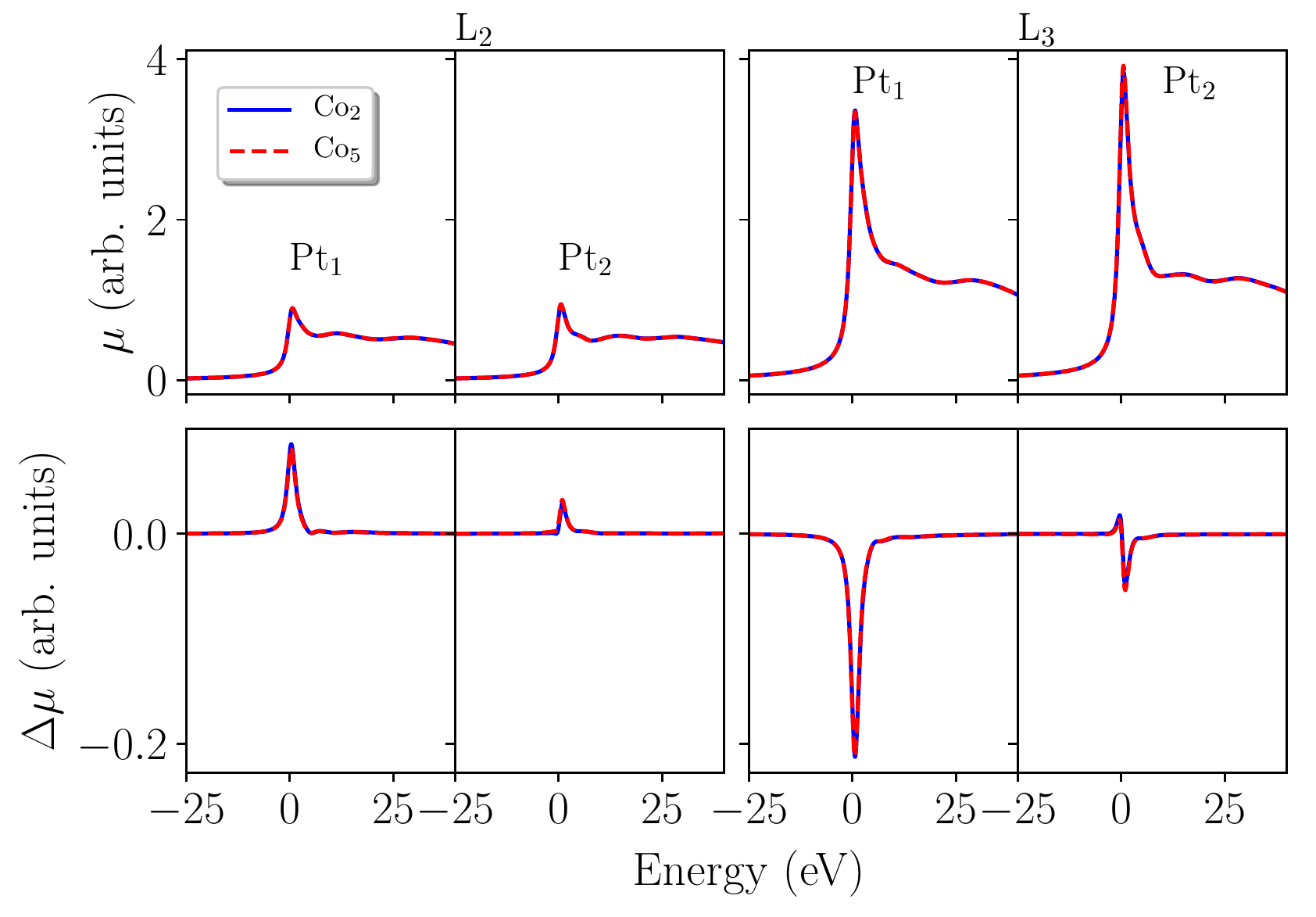}
\caption{Calculated layer resolved XAS (top panel),
 $\mu$ and XMCD (bottom panel),
  $\Delta \mu$ spectra  at the L$_{2}$ and L$_{3}$ edges for the Pt layers in 
 Pd(001)/Co$_{2}$/Pt$_{2}$ (solid line) and Pd(001)/Co$_{5}$/Pt$_{2}$ (dashed line).}
\label{fig_xas-xmcd_Pt2}
\end{figure}

As discussed in the literature \cite{PhysRevLett.120.157203}, the above-mentioned
 field-induced changes of the electronic structure and magnetic properties
can be probed in a detailed way 
using  XAS/XMCD spectroscopy. Focusing here on the
magnetic properties of the  Pt layers, the absorption spectra at the Pt $L_2$ and
$L_3$ edges have been calculated both for Pd(001)/Co$_{2}$/Pt$_{2}$ and
Pd(001)/Co$_{5}$/Pt$_{2}$.
 The XAS and XMCD spectra without the influence
 of an external electric field
are given in 
Fig.\ \ref{fig_xas-xmcd_Pt2}, 
showing only tiny differences between the two systems.
%
%
%

\begin{figure}[htb]
\centering
\includegraphics[width=1.00\columnwidth]{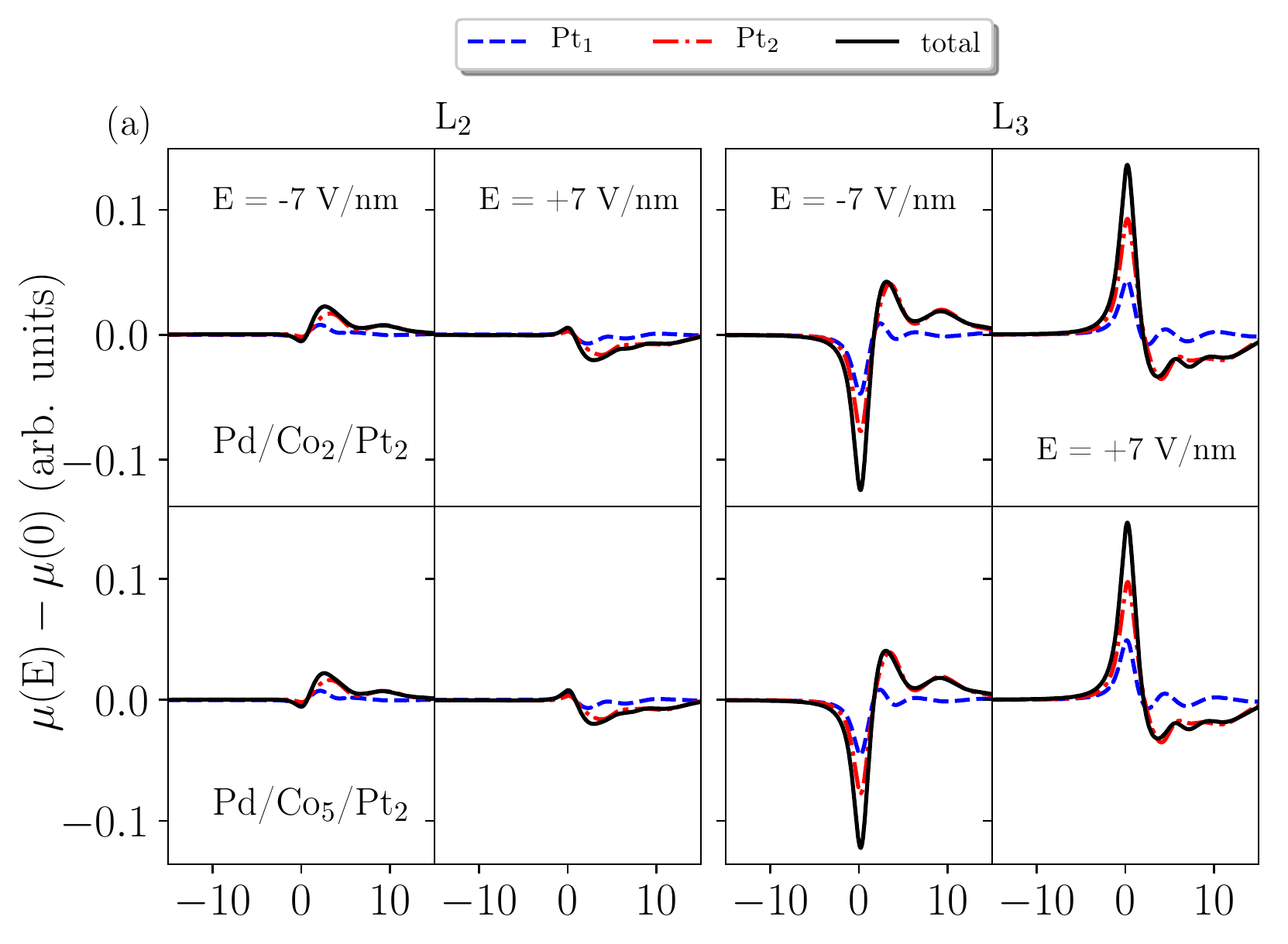}\\
\includegraphics[width=1.00\columnwidth]{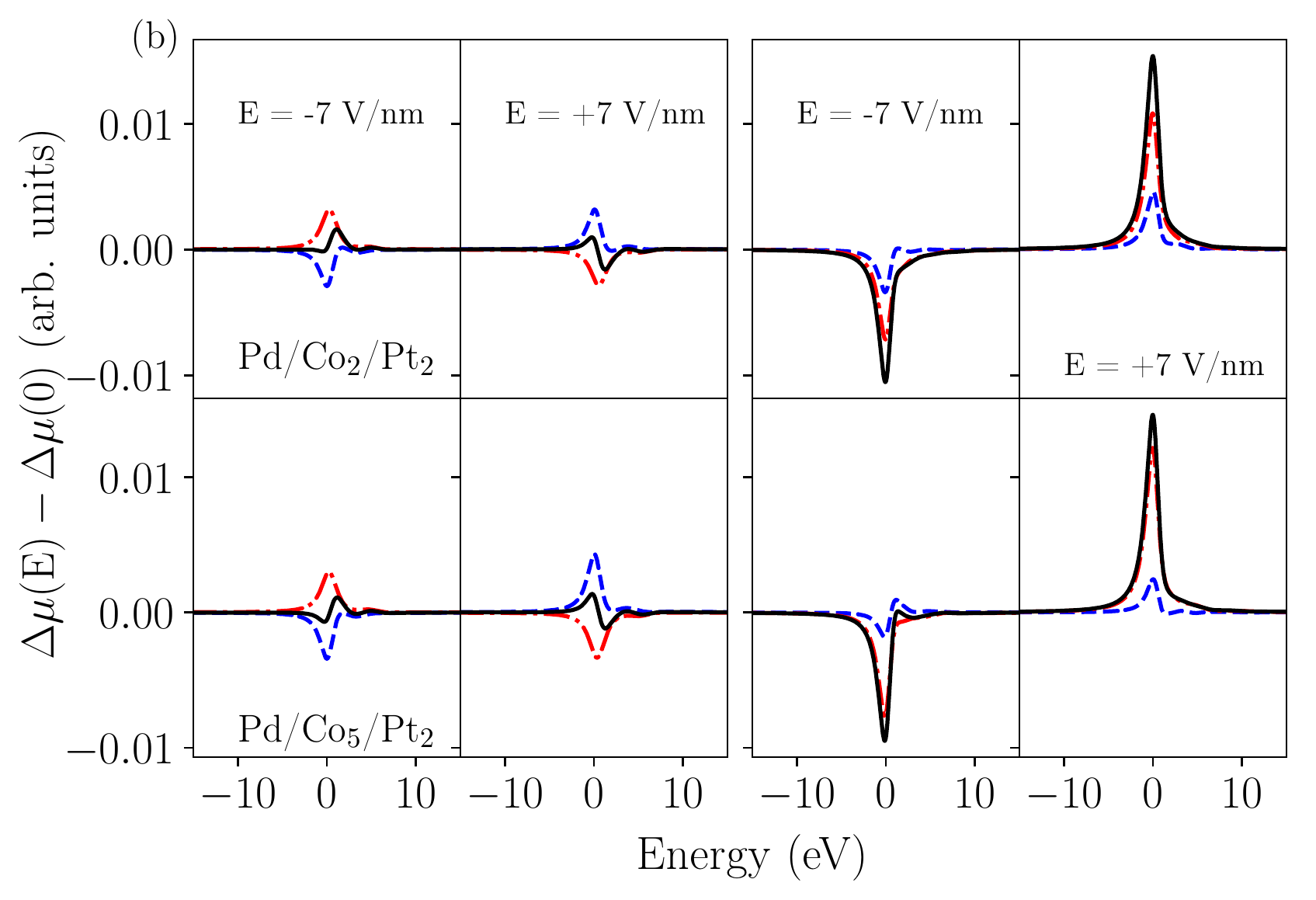}\\
\caption{Difference of the calculated layer resolved XAS (a) and XMCD  (b) spectra of Pt layers between in absence and presence of an electric field,
 $\mu(\pm E) -\mu(0)$,
 $\Delta \mu(\pm E) -\Delta \mu(0)$  around L$_{2}$ and L$_{3}$ edges in case of Pd(001)/Co$_{2}$/Pt$_{2}$ and Pd(001)/Co$_{5}$/Pt$_{2}$  systems.}
\label{fig_xas-xmcd-E_Pt2}
\end{figure}

The modification of the XAS spectra 
for these systems by
the electric field   $\pm E$, are represented
 in 
 Fig.\ \ref{fig_xas-xmcd-E_Pt2}, 
 showing the field-induced changes,
$\mu(\pm E)-\mu(0)$, of the total XAS and of XMCD signals,  $\Delta \mu(\pm E)-\Delta \mu(0)$ for the Pt 
layers in Pd(001)/Co$_{2}$/Pt$_{2}$  and  Pd(001)/Co$_{2}$/Pt$_{4}$.
Although only  tiny differences are found for both systems,
one can see that the changes are most pronounced
at the L$_{3}$ edge in both cases.
This finding is in line with the
previously reported experimental results  \cite{PhysRevLett.120.157203}
and can be explained as follows.
As an electric field will in particular shift 
electronic states below or above the Fermi level,
 pronounced field induced changes have to be expected
first of all   at the absorption edges
of  the spectra.
As the Pt d-states in that energy region 
have primarily $d_{5/2}$-character
and as the L$_{3}$ and L$_{2}$ spectra are dominated 
by their $d_{5/2}$- and $d_{3/2}$-contributions, 
respectively, it follows that an
electric field  has a much stronger impact for the 
L$_{3}$ than for the  L$_{2}$ spectrum.  

The modifications of XAS and XMCD spectra  
depend directly on the
direction of the electric field as this determines the
direction of the field-induced shift of  
the electronic states with respect to the 
Fermi energy. 
Due to the screening of the electric field 
with increasing distance from the
surface, the field-induced changes of the XAS and XMCD signals
are most pronounced for
 the surface Pt layer and decrease towards the
interface. This behavior can   be seen clearly in 
Fig.\ \ref{fig_xas-xmcd-E_Pt2} 
showing
the layer resolved results for 
Pd(001)/Co$_{2}$/Pt$_{2}$. 
The same trend
is found also for Pd(001)/Co$_{5}$/Pt$_{2}$.

\subsection{Variation of the thickness of the Pt layer }
\begin{figure}[htb]
\centering
\includegraphics[width=1.00\columnwidth]{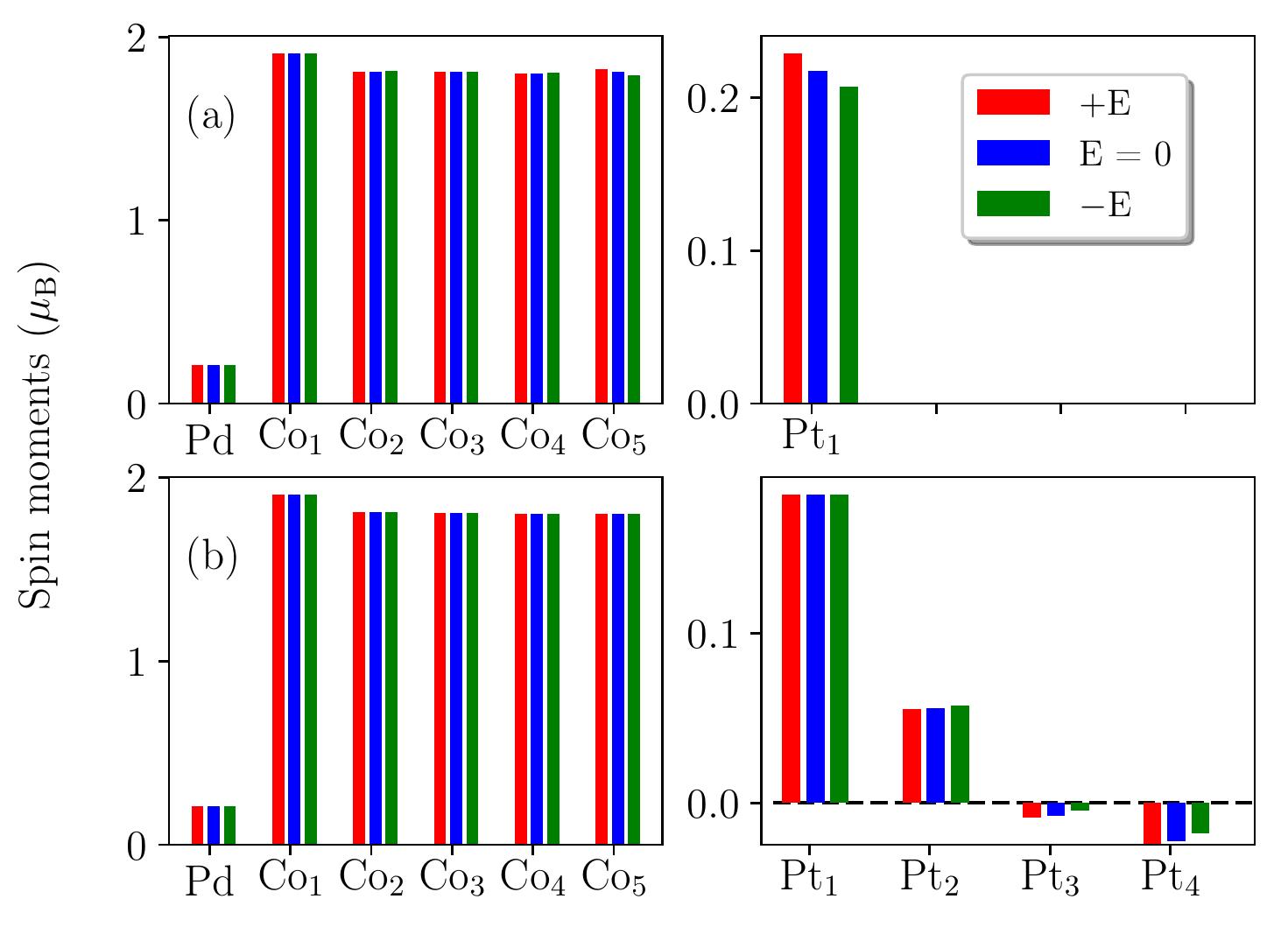}
\includegraphics[width=1.00\columnwidth]{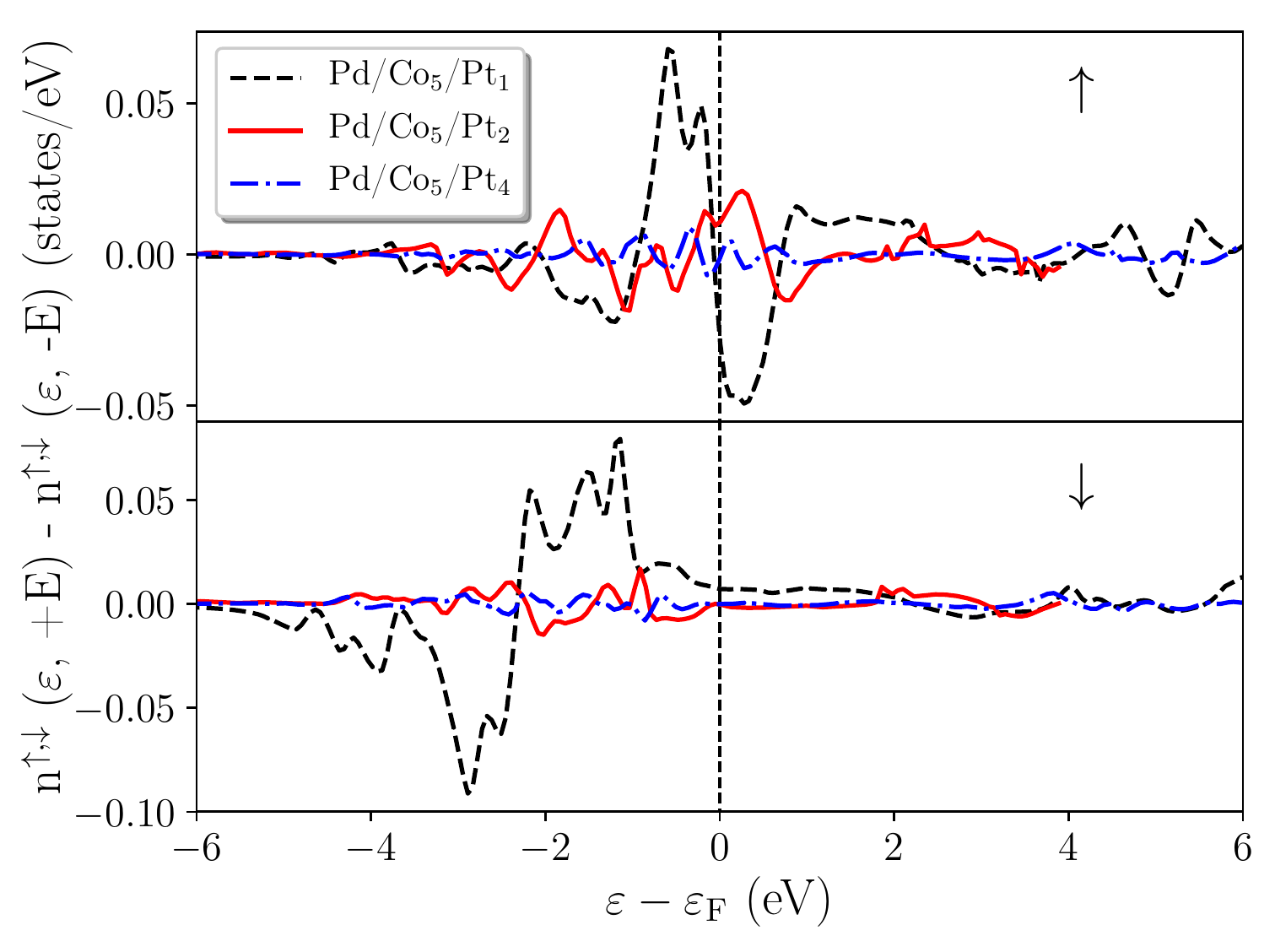}%
\caption{Top panel: 
Calculated 
layer resolved 
magnetic spin moments of  the individual atomic layers in 
Pd(001)/Co$_{5}$/Pt$_{1}$ (a) and Pd(001)/Co$_{5}$/Pt$_{4}$ without electric field and in the presence of a positive and negative electric field. Bottom panel: Calculated difference for the minority and majority density of states  
  of the interface Co layer 
  between positive and negative electric field
for  different thicknesses of the capping Pt layer.}
\label{fig_diff_DOS_Co5}
\end{figure}

In order to investigate how the electric field effect changes with an increasing thickness of the Pt film, calculations have been performed for the Pd(001)/Co$_{5}$/Pt$_{m}$ system, where the thickness of the capping Pt layer was varied between $m=1$ and $m=4$.
The top panel of 
Fig.\ \ref{fig_diff_DOS_Co5} 
shows the calculated spin moments of the individual layers in  
Pd(001)/Co$_{5}$/Pt$_{1}$ (a) and Pd(001)/Co$_{5}$/Pt$_{4}$ 
 for various electric fields.
%
%
In the case of Pd(001)/Co$_{5}$/Pt$_{1}$  the Pt spin moment is 
$m_{\rm Pt_{1}}=0.22\, \mu_{B}$. For Pd(001)/Co$_{5}$/Pt$_{4}$ 
 the induced moment of the Pt layers significantly decreases away from the Co interface while the Pt layer at the Co interface possesses the largest induced moment with 
$m_{\rm Pt_{1}}=0.18\, \mu_{B}$
 coupled ferromagnetically to that of Co. The spin magnetic moment of the next Pt layer is 
$m_{\rm Pt_{2}}=0.06\, \mu_{B}$ and is also ferromagnetically aligned to the Co moments. The remaining very small magnetic moments for the third and fourth Pt layers are antiferromagnetically oriented with respect to the Co layers.

\begin{figure}[htb]
\centering
\includegraphics[width=1.00\columnwidth]{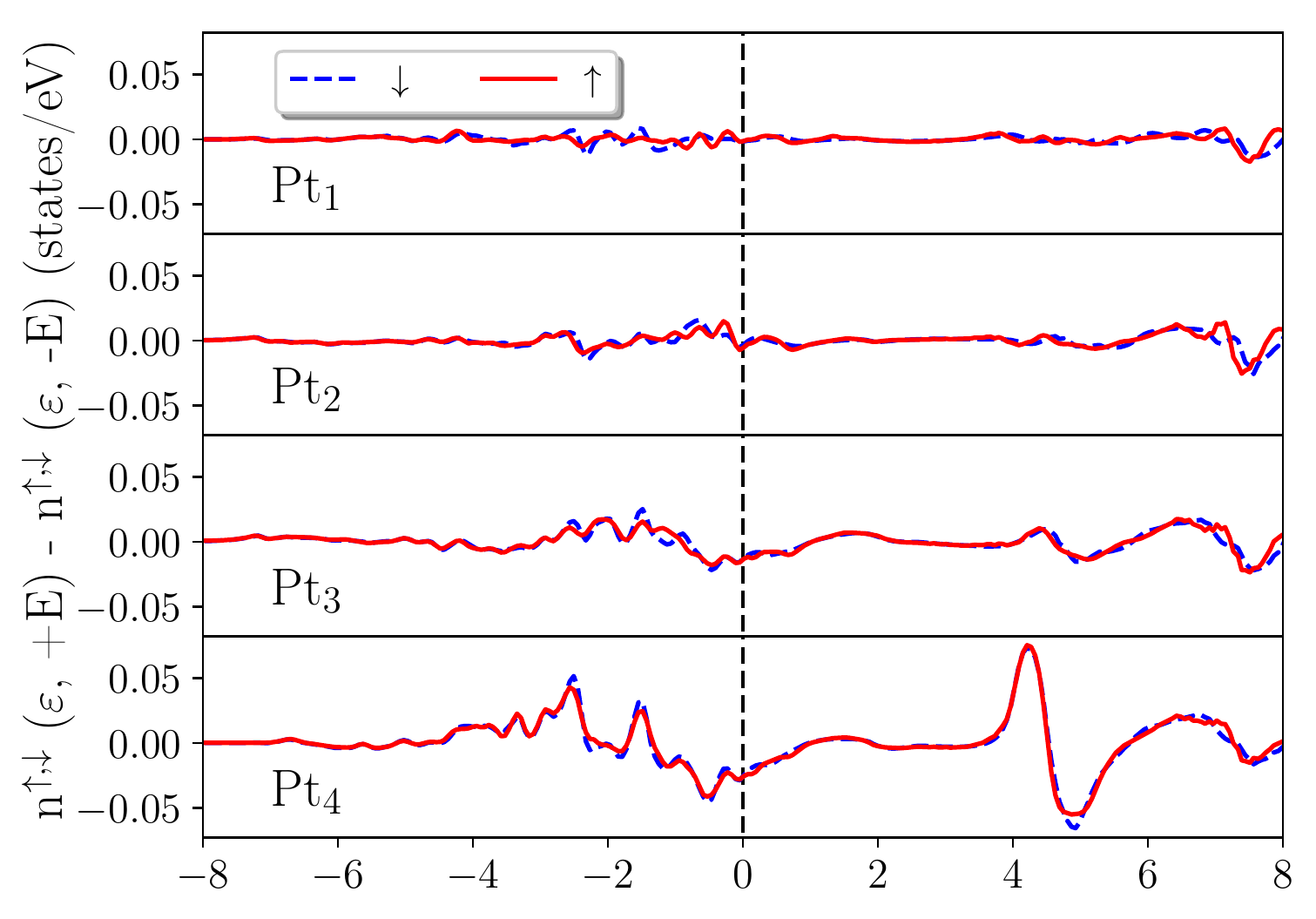}\\
\includegraphics[width=1.00\columnwidth]{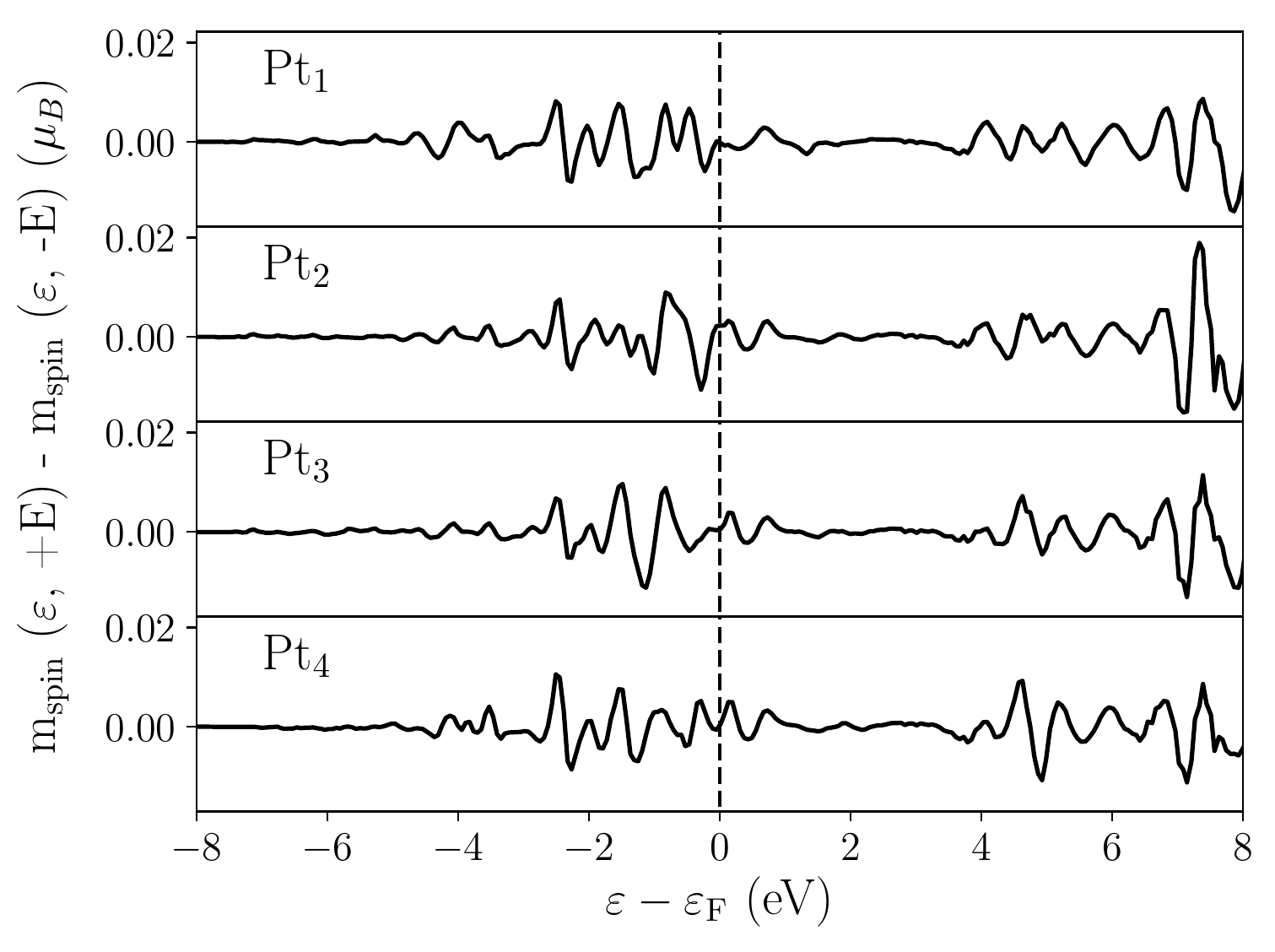}
\caption{Top panel: Electric-field-induced change of the density of states
in the different Pt layers in  Pd(001)/Co$_{5}$/Pt$_{4}$. 
Bottom panel: Electric-field-induced change of the difference $\rm m_{\rm spin} = n^\uparrow(\varepsilon) - n^\downarrow(\varepsilon)$ of majority- and minority DOS.  }
\label{fig_DOS_Pt}
\end{figure}

The Pt spin-magnetic moment in  Pd(001)/Co$_{5}$/Pt$_{1}$ 
 increases in case of a positive electric
field ($+ E$) while it 
decreases for $- E$ with the value
$0.23 \, \mu_{B}$ and 
$0.21 \, \mu_{B}$, respectively. 
This dependency on the electric field is opposite to that of 
Pd(001)/Co$_{n}$/Pt$_{2}$  
shown  in 
Fig.\ \ref{fig_momPt}
and can be attributed to the screening of the electric field that gets more important for an increasing thickness of the Pt film. In particular, one can see rather strong field-induced changes of the spin magnetic moments of the interface and next-to-interface Co layers in  Pd(001)/Co$_{5}$/Pt$_{1}$.
 In this case the spin moment of the Pt layer follows the changes of the spin moment of Co at the Co/Pt interface. Obviously, the impact of the electric field on the Co spin moment significantly decreases with increasing of the thickness of the Pt film.
The bottom panel in 
Fig.\ \ref{fig_diff_DOS_Co5} 
shows the difference in the density of states
for majority and minority spins
of the topmost Co layer (Co$_{5}$) obtained for different electric fields. These electric fields induced changes in the DOS are decreasing for the Co layers when the thickness of Pt film increases.
This screening effect is also seen in the bottom panel of 
Fig.\ \ref{fig_DOS_Pt} 
which represents the field induced DOS and the spin density changes in the different Pt layers
  of Pd(001)/Co$_{5}$/Pt$_{4}$. 
%
%
%
One can see that the most pronounced DOS changes due to an
applied electric field occur in the surface Pt layer, while the DOS modification in the deeper Pt layer and at the Pt/Co interface is rather weak (see top panel of
 Fig.\ \ref{fig_DOS_Pt}).
Despite this trend, the bottom panel of 
Fig.\ \ref{fig_DOS_Pt} 
shows that the field induced changes of the spin polarization (i.e.\ m$_{spin}(\varepsilon) = n^\uparrow(\varepsilon) - n^\downarrow(\varepsilon)$ have the same order of magnitude for all Pt layers. This can be attributed to the strongest field effect for the surface Pt layer on the one hand side and the strongest proximity induced spin moment at the interface on other hand side. 
  
Thus, one can conclude that depending on the thickness of the Pt layer, the field dependent changes of the induced spin moment can be dominated by different  mechanisms associated either with field induced changes of the electronic structure and magnetic moments in the ferromagnetic  sub-surface 
or in the non-magnetic surface parts
 of the  system.

\begin{figure}[htb]
\centering
\includegraphics[width=1.00\columnwidth]{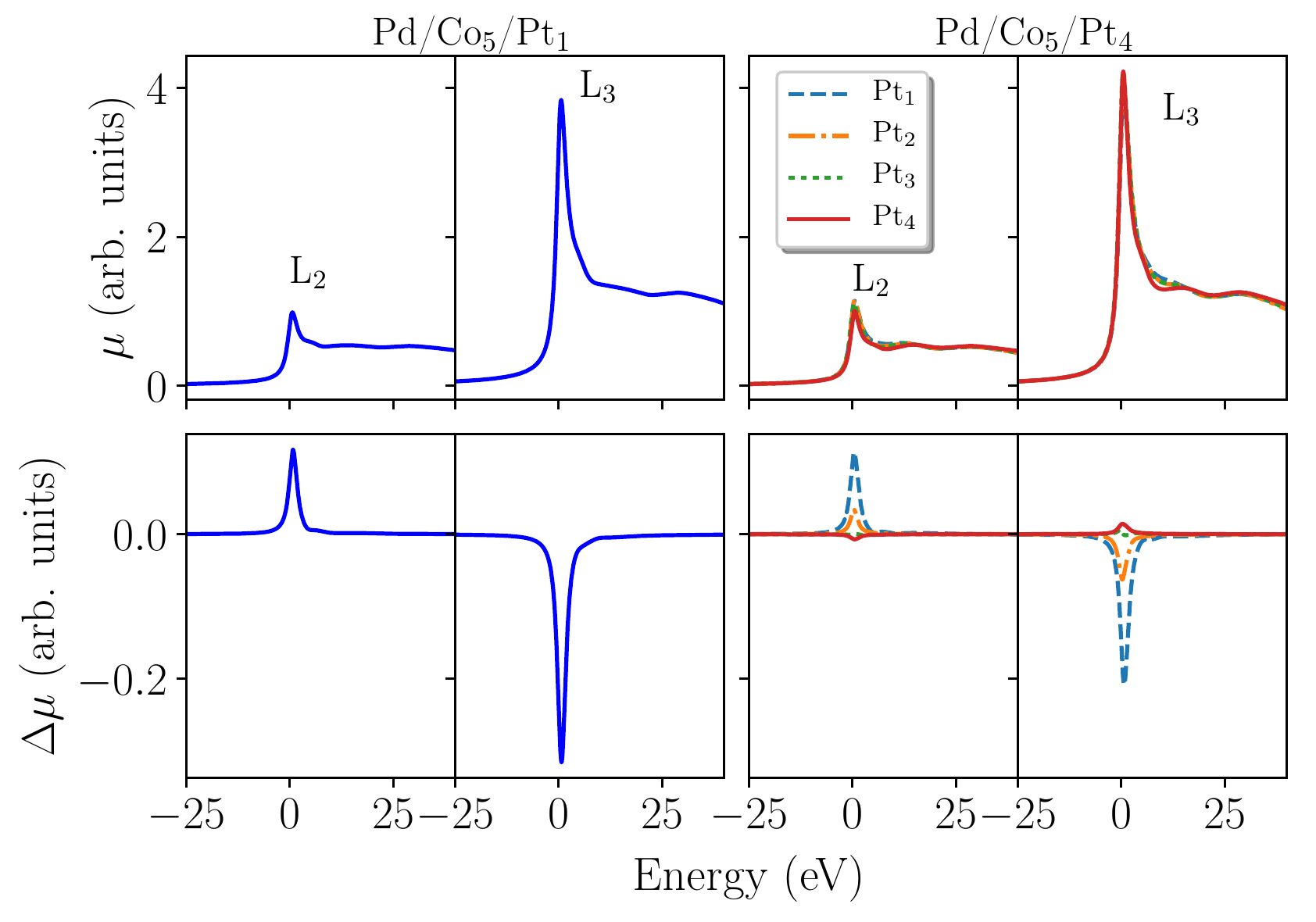}
\caption{Calculated layer resolved 
XAS (top panel)  $\mu$ and
XMCD (bottom panel) $\Delta \mu$
  spectra at the 
L$_{2}$- and L$_{3}$-edges for the Pt layers in 
 Pd(001)/Co$_{5}$/Pt$_{1}$ (left) and Pd(001)/Co$_{5}$/Pt$_{4}$ (right). } 
\label{fig_xas-xmcd_Pt1-Pt4}
\end{figure}

Next, we discuss the influence of the electric field 
on the XAS and XMCD
spectra at the L$_{2}$- and L$_{3}$-edges of Pt in  Pd(001)/Co$_{5}$/Pt$_{n}$
 with $n = 1$ and $4$ and its
dependence  on the
thickness $n$ of the Pt film.
First, we consider in 
Fig.\ \ref{fig_xas-xmcd_Pt1-Pt4} (top panel) 
the layer resolved XAS spectra calculated without including an external electric
field.   
%
%
%
One can see a weak dependence of the XAS spectra  on the position of Pt layer in the 
Pd(001)/Co$_{5}$/Pt$_{4}$ system. However,
the XMCD spectra shown in 
Fig.\ \ref{fig_xas-xmcd_Pt1-Pt4} (bottom panel), 
exhibit a  rather pronounced decrease, when going from the interface
(Pt$_{1}$)  to the surface (Pt$_{4}$) layer.
Note that the  XMCD 
signal  of the surface Pt layer even changes sign in line with the
sign change for
 the induced spin moment in this layer 
(see discussion above). 
The strongest XMCD signal occurs for the interface Pt layer reflecting the largest induced spin moment due to proximity to the ferromagnetically ordered Co layers.

\begin{figure}[htb]
\centering
\includegraphics[width=1.00\columnwidth]{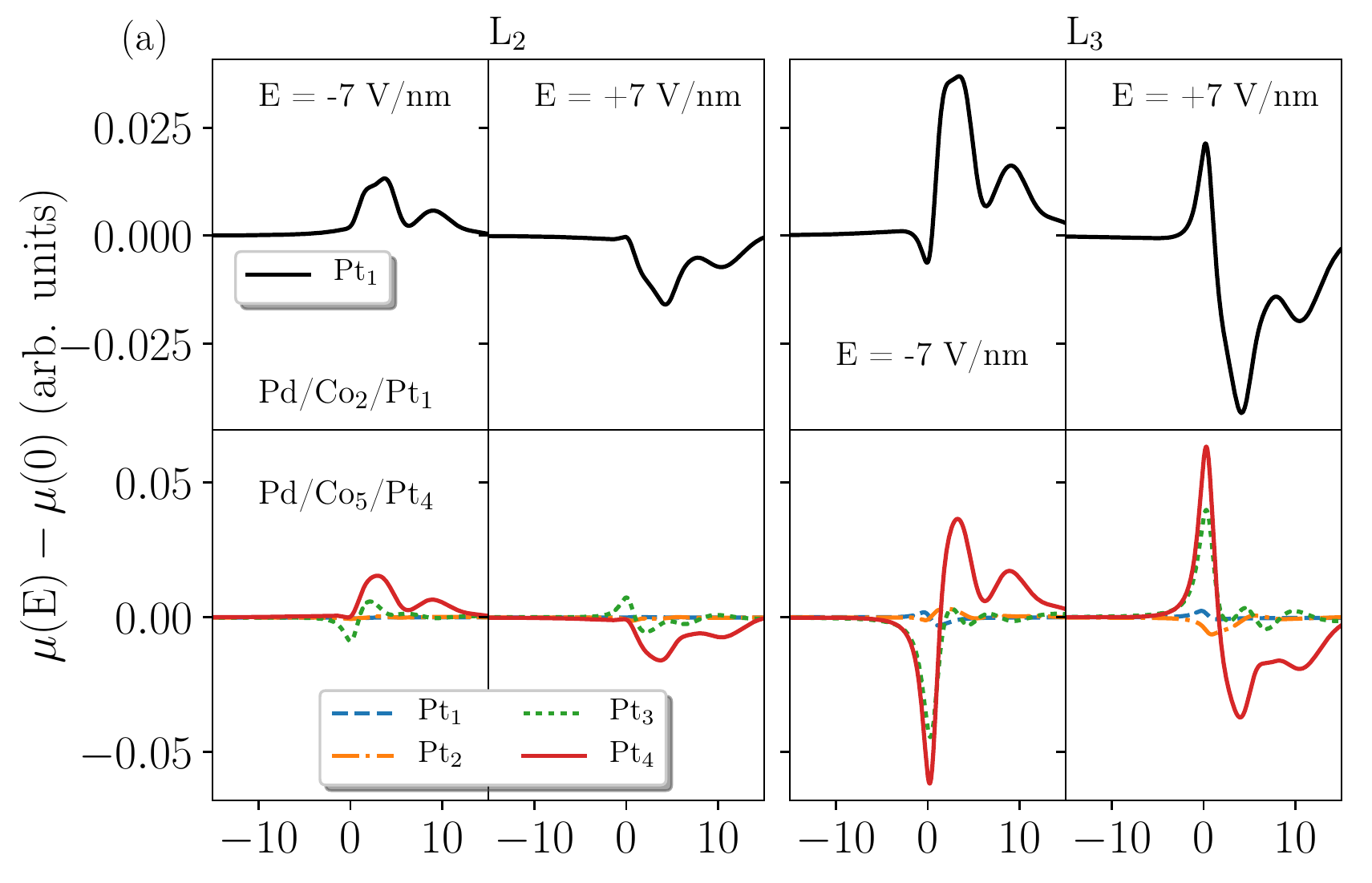}\\
\includegraphics[width=1.00\columnwidth]{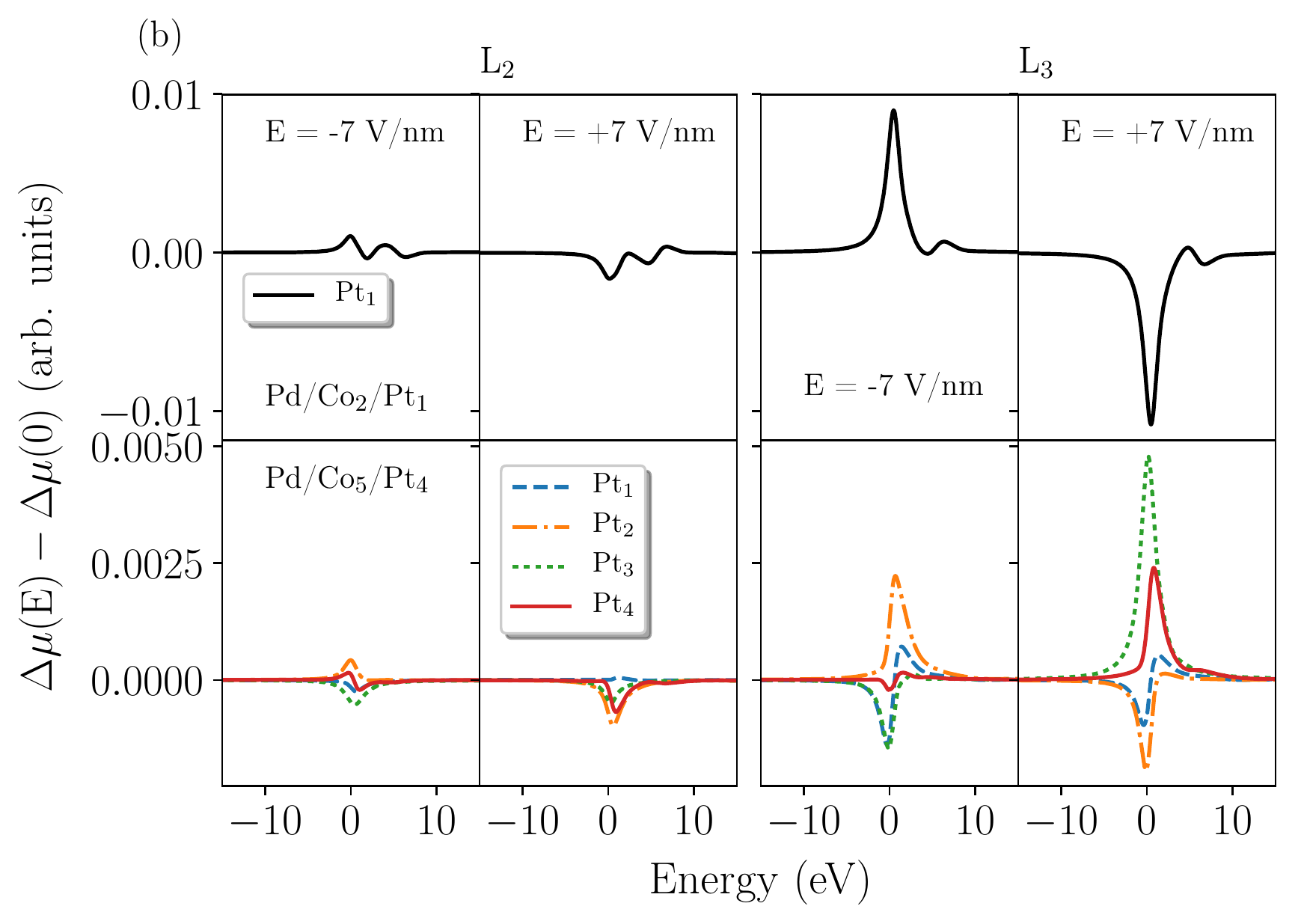} 
\caption{Electric field induced change in
the XAS (a)
  and XMCD (b) spectra  at  the 
L$_{2}$ and L$_{3}$ edges 
  for Pt in  
Pd(001)/Co$_{5}$/Pt$_{1}$ and 
Pd(001)/Co$_{5}$/Pt$_{4}$.}
\label{fig_diff_xmcd-E_Pt1-Pt4}
\end{figure}

The changes of the XAS and XMCD spectra of Pt in
Pd(001)/Co$_{5}$/Pt$_{1}$ and 
Pd(001)/Co$_{5}$/Pt$_{4}$
that are caused by the
applied electric field   are presented in
 Figs.\ \ref{fig_diff_xmcd-E_Pt1-Pt4}, (a) and (b), respectively.
%
%
Similar to  the systems with 
$2$~ML of Pt, 
one finds that the change of the 
spectra reverses its sign if the orientation of the electric field is reversed.
In addition, one can see 
the asymmetry of these
modifications with respect to a change in the orientation of the electric field. 
The most pronounced changes occur for the Pt L$_{3}$ edge signal,
similar to the results obtained
for Pd(001)/Co$_{2}$/Pt$_{2}$.  
The intensities of the layer-resolved changes of the  XAS spectra for Pt in
Pd(001)/Co$_{5}$/Pt$_{4}$
gradually decrease when going  from the  Pt surface  to the interface layer. As discussed above, this
 can be attributed to the screening of the electric field in the surface region.
However, the XMCD spectra change
non-monotonously towards the interface 
layer as a consequence of
 a competition of the 
decreasing electric field strength
and the increasing impact of the neighboring Co layers controlling 
the Pt spin magnetic moment via the proximity effect.

\begin{figure}[htb]
\centering
\includegraphics[width=1.00\columnwidth]{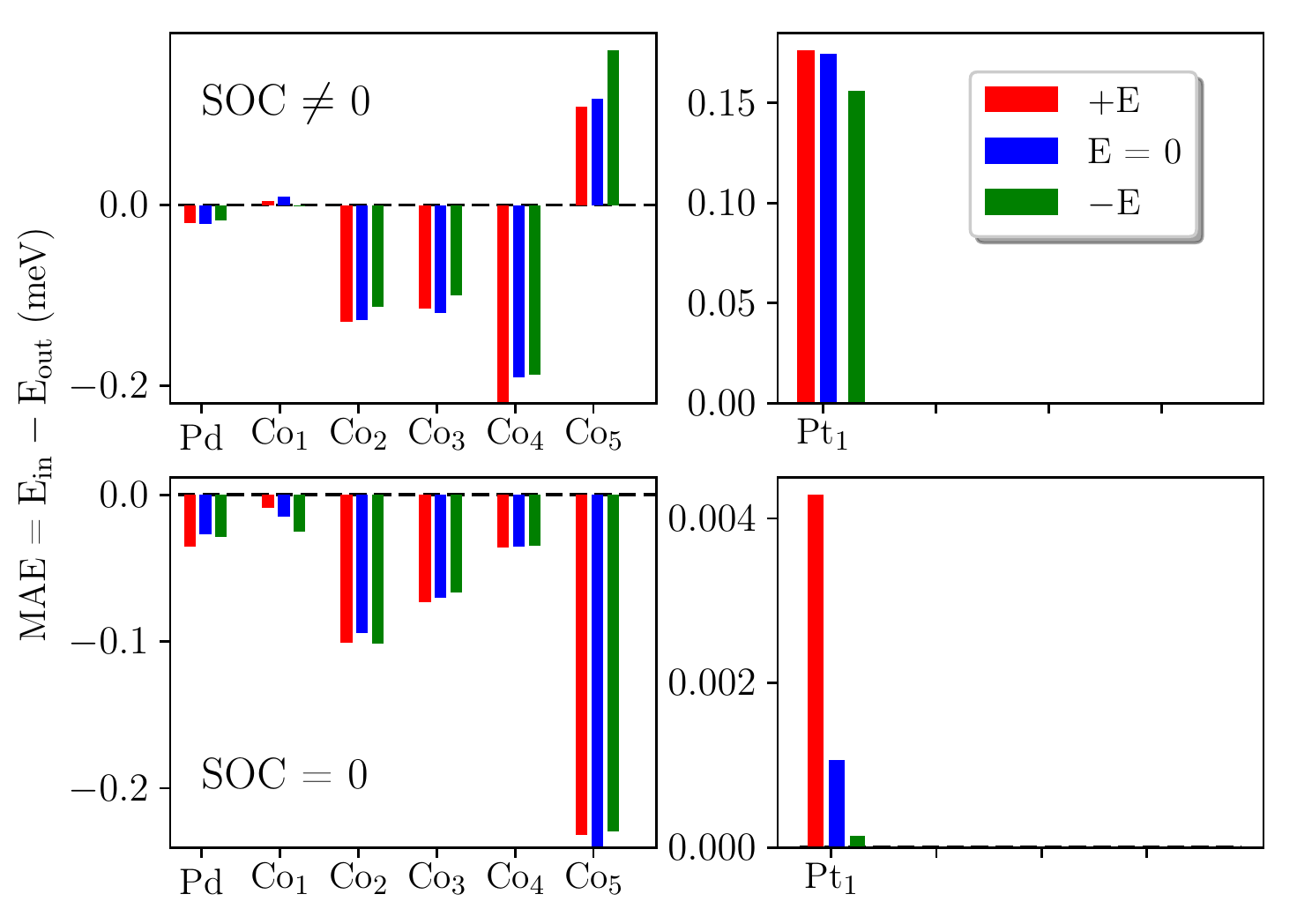} 
\includegraphics[width=1.00\columnwidth]{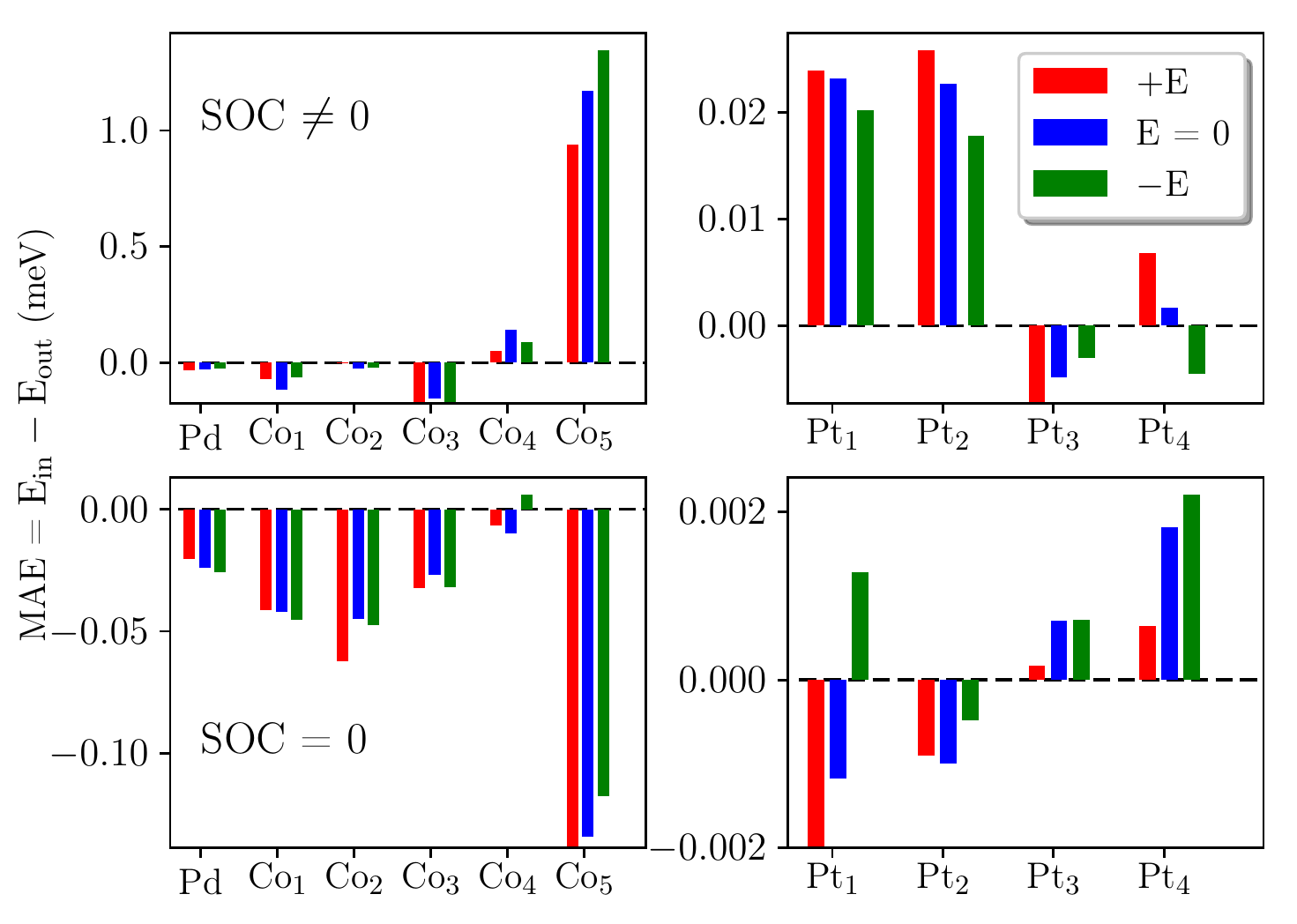}
\caption{Calculated layer-resolved magnetic anisotropy energy (MAE) of
Pd(001)/Co$_{5}$/Pt$_{1}$ (top panel) and
Pd(001)/Co$_{5}$/Pt$_{4}$  (bottom panel) 
for no electric field present
 as well as for
 an applied electric field with positive and negative sign, respectively.  
The results have been obtained from 
fully relativistic calculations 
(SOC $\neq$ 0) and calculations
with the strength  of the spin-orbit
 coupling   in the Pt layers
set to zero  (SOC = 0).}
\label{fig_MAE-E_Pt1-Pt4}
\end{figure}

In addition to the impact of an 
electric field on the XMCD spectra, 
its influence on the layer
resolved MAE was  investigated. 
Figure \ref{fig_MAE-E_Pt1-Pt4}
shows the layer resolved MAE of  Pd(001)/Co$_{5}$/Pt$_{1}$ and
Pd(001)/Co$_{5}$/Pt$_{4}$, respectively,
 for no electric field present
 as well as for
 an applied electric field with positive and negative sign, respectively. 
%
%
%
The definition for the MAE used implies an
out-of-plane and in-plane anisotropy
for a positive or
negative, respectively, sign of  the MAE.

In an earlier study  \cite{PhysRevLett.102.187201}
it was already found that an  electric field
may strongly affect the magneto-crystalline anisotropy of free-standing transition metal  mono layers,
 as a 
result of the distortion of the  electronic structure
by the applied electric field. 
The authors point out that in these systems the 
electric field breaks the z-reflection symmetry, 
lifting that way a degeneration in the 
cross-points of the energy bands via the
field-induced hybridization between the $d$ and $p$ orbitals that contribute to these states.

\begin{figure}[htb]
\centering
\includegraphics[width=1.00\columnwidth]{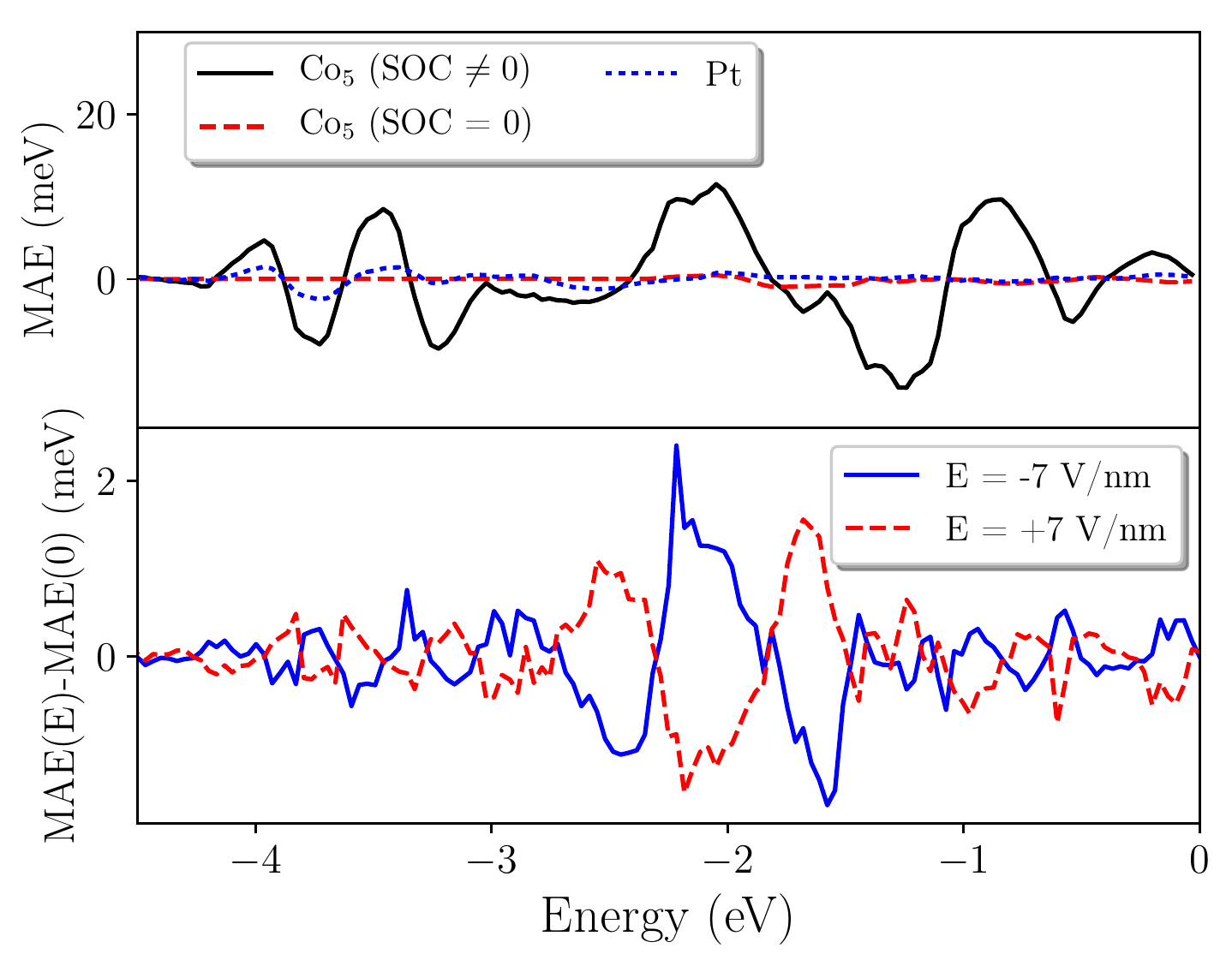}
\caption{Top panel: Contribution to the magnetic anisotropy energy from the Co$_{5}$ and Pt layers
  (without SOC scaling and with SOC = 0 in the Pt layers), represented as a
 function of occupation of energy bands in case of Pd(001)/Co$_{5}$/Pt$_{1}$
  system. Bottom panel: Field-induced changes of the contribution to the magnetic
  anisotropy energy from the Co$_{5}$ layer, MAE($\pm E$)- MAE(0). }
\label{fig_diff_Co5_MAE-E_Pt1}
\end{figure}

\begin{figure}[htb]
\centering
\includegraphics[width=1.00\columnwidth]{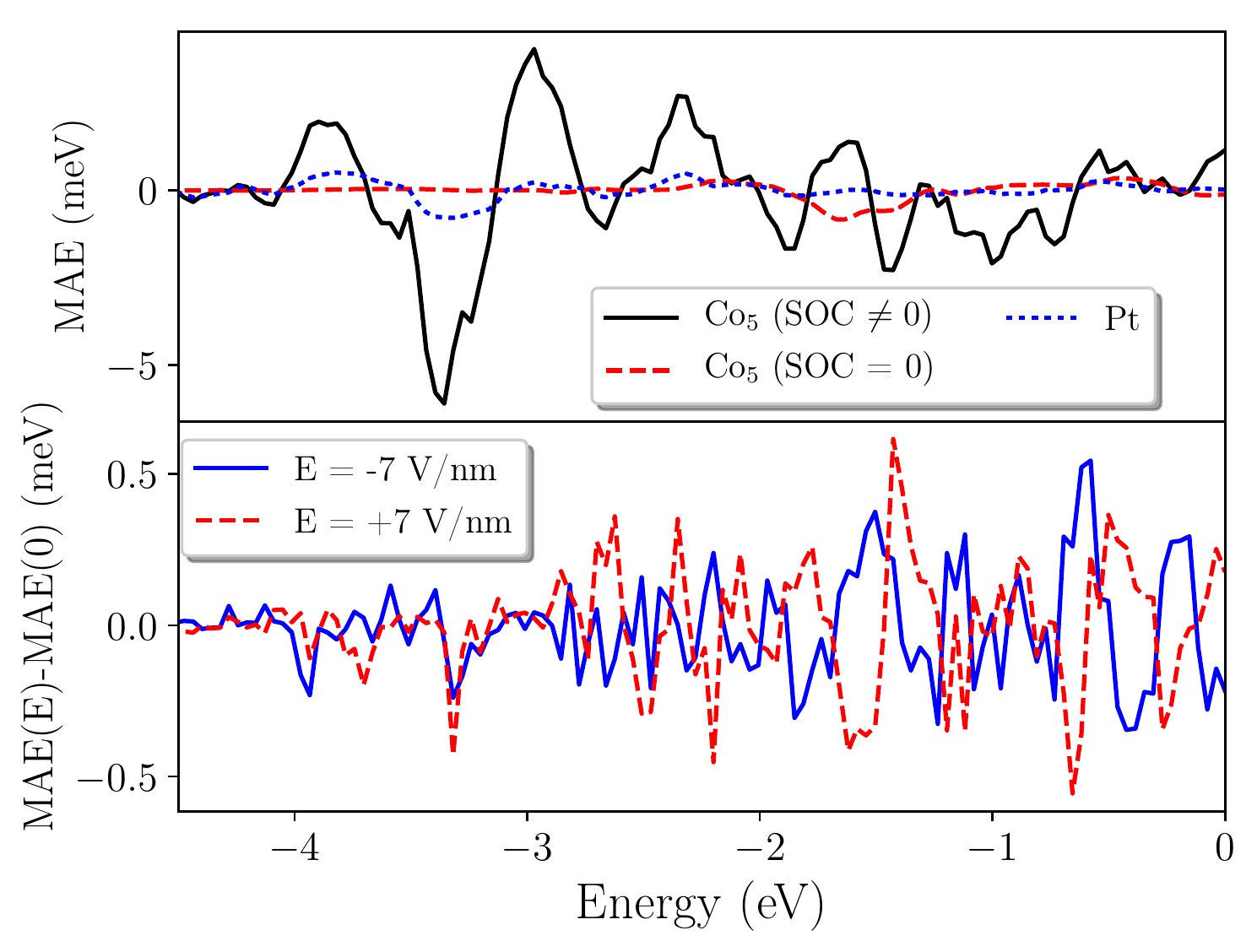}
\caption{Top panel: Contribution to the magnetic anisotropy energy from the Co$_{5}$ and Pt layers
  (without SOC scaling and with SOC = 0 in the Pt layers), represented as a
 function of occupation of energy bands in case of Pd(001)/Co$_{5}$/Pt$_{4}$
  system. Bottom panel: Field-induced changes of the contribution to the magnetic
  anisotropy energy from the Co$_{5}$ layer, MAE($\pm E$)- MAE(0). }
\label{fig_diff_Co5_MAE-E_Pt4}
\end{figure}

As Fig.\ \ref{fig_MAE-E_Pt1-Pt4} 
illustrates,
an electric field also strongly 
modifies the MAE for the systems 
considered here  despite the lack of 
z-reflection symmetry for the field-free case.
As a reference, the figure also shows
the  total and layer resolved  MAE for
Pd(001)/Co$_{5}$/Pt$_{1}$ 
and Pd(001)/Co$_{5}$/Pt$_{4}$  
for zero-field conditions.
As one can see,  Pd(001)/Co$_{5}$/Pt$_{1}$
has a total  MAE corresponding to an
in-plane anisotropy with 
dominating contributions from the Co layers
in the middle of the Co film, 
while the positive contribution from the
interface Co/Pt layer is rather small.
Pd(001)/Co$_{5}$/Pt$_{4}$, on the other hand, 
 has out-of-plane
anisotropy with its 
 MAE  dominated by 
the contribution from the 
Co layer at the Co/Pt interface.
It should be noted that a strong 
dependence of the MAE on the thickness 
of the over layer
was discussed already previously 
for several systems
\cite{Beauvillain_JAP_1994, PhysRevLett.77.1805}. 
Figures \ref{fig_diff_Co5_MAE-E_Pt1}
 and
\ref{fig_diff_Co5_MAE-E_Pt4} 
illustrate the contributions to the MAE  from the
interface Co and Pt layers, 
represented as a function of the occupation of
valence band realized by artificially 
varying the Fermi energy.
%
%
%
%
%
These relations have a 
non-monotonous behavior with 
extreme values always 
occurring when the  Fermi energy is passing
through a SOC-induced avoided crossing of the energy bands. 
One can see that for both systems, the amplitude of 
the contribution to the
MCA  associated with the  Co interface layer is much 
larger
than that for the  Pt  interface layer. 
 However, accidentally, these values can be
close to each other at a  certain  occupation,
as it is the case for
 Pd(001)/Co$_{5}$/Pt$_{1}$ at the proper  Fermi energy.
 It should be noted
  that the MAE of  materials composed of 
   magnetic- and
heavy-element components 
is determined essentially by the spin-dependent
hybridization of their electron orbitals 
as well as by  the SOC of the heavy element 
(see for example 
Ref.\ \onlinecite{ASE+07,SMME08,OKS+17}).
As the  Pd(001)/Co$_{5}$ substrate is 
common to both systems considered, the 
difference in the MAE has to be
 attributed to the details of the
electronic structure of these systems 
associated with a different
thickness of the Pt surface film. 
As a result, switching the SOC on the Pt atoms
artificially off
leads for the Pt film
 to a  strong weakening of
 the dependence of its electronic structure
 on the magnetization
direction  and in turn of the MAE.
This is seen in 
Figs.\  \ref{fig_diff_Co5_MAE-E_Pt1} 
and     \ref{fig_diff_Co5_MAE-E_Pt4}  
indicating that the contribution
of the Co interface layer to MAE
 drops down by about one order of magnitude 
 when the SOC is switched  off for Pt.
 For this situation, 
 the layer resolved MAE for both systems 
is rather similar 
(see Fig.\ \ref{fig_MAE-E_Pt1-Pt4}). 
The difference between the results
for the two systems  can be attributed
to some extent 
to a different hybridization of the 
Co and Pt related  electronic states, which is
obviously dependent on the thickness of the Pt film.
Based on these results, 
 one can expect that  the field
induced changes to the MAE should be 
 associated first of all to the
influence of the electric field on the Pt related electronic states.

Analysing  the field-induced changes of the total and layer resolved
MAE of 
Pd(001)/Co$_{5}$/Pt$_{1}$ and Pd(001)/Co$_{5}$/Pt$_{4}$,
the most pronounced changes are found
 at the interface, although the changes  for the other layers are not negligible.
Despite the pronounced 
screening effect in  case of 
Pd(001)/Co$_{5}$/Pt$_{4}$ with  4~MLs of Pt, the 
field-induced change of the  
Co interface contribution to the MAE is
significant, indicating a key role of the electronic structure changes
occurring in the Pt film due to the electric field.   
The field induced changes of the MAE in the systems with $1$ and $4$ Pt monolayers
are associated  
primarily with the Co/Pt interface contribution.
Corresponding results for the interface layers are 
 plotted in Figs.\ \ref{fig_diff_Co5_MAE-E_Pt1} and
 \ref{fig_diff_Co5_MAE-E_Pt4}, respectively, as a function of the
 occupation. From these figures one can see that for most of the 
occupation numbers the MAE changes have opposite sign for the opposite orientation when the  electric field
is reversed.
The origin of these changes of the MAE can be attributed first of all to
the modification of the electronic structure of 
the Pt film, i.e. the electric field controlled hybridization of
the $d$  and $p$ orbitals, as discussed in 
Ref.\ \onlinecite{PhysRevLett.102.187201}.

\section{Conclusions \label{sec4}}
In conclusion, in this work we examined  the
influence of an  electric field effect 
on the magnetic properties 
in the Pt layer of Pd(001)/Co$_{n}$/Pt$_{m}$ thin film structures by performing
 first-principles calculations. 
For this purpose, a homogeneous
 external electric field was modeled by a 
 charged  plate in front of the surface. 
 From the self-consistent calculations, we determined the spin magnetic moment and XMCD spectra in the presence of an electric field. We found that 
 in case of 
 Pd(001)/Co$_{2}$/Pt$_{2}$ and 
 Pd(001)/Co$_{5}$/Pt$_{2}$
 that 
  the spin-magnetic moments are varying 
    independently from the number of  Co layers 
  roughly  quadratically as a function
   of the electric field strength.
   An inspection of the angular momentum 
   resolved DOS reveals that 
    the electric field slightly shifts the $s$ and $p$ states around the Fermi level.
     From the calculated XMCD spectra, 
     it was found that the electric field 
   has its major impact 
   for the  L$_{3}$ edge spectra. 
   We also investigated 
dependency of the    electric field effect 
on the  thicknesses of the Pt layer. 
In case of  
Pd(001)/Co$_{5}$/Pt$_{1}$ as well as  
Pd(001)/Co$_{5}$/Pt$_{4}$, the electric field
induced 
change of the XMCD spectra is most
 significant for the L$_{3}$ edge, independent on the thickness of the Pt capping layer.
 In addition, the layer dependent MAE 
 and its dependency on an electric field 
 was examined. It was found that the electric field strongly modifies the MAE.
 It turned out in particular 
 that this change is still considerable for
  deeper lying layers.

\begin{acknowledgments}
This work was supported by the Deutsche Forschungsgemeinschaft
grant: DFG EB 154/35.
\end{acknowledgments}

\bibliography{references.bib}

\end{document}